\documentclass[pra,amsmath,aps,10pt,superscriptaddress,letterpaper,tightenlines]{revtex4}
\usepackage{amsmath}
\usepackage{bm}
\usepackage{stmaryrd}
\usepackage{amssymb}
\usepackage{graphicx}
\usepackage{textcomp}
\usepackage{calrsfs}
\usepackage{yfonts}
\usepackage{color}
\newcommand{\p}{\partial}
 \begin{document}
\title{On the localized quantum oscillators in a common heat bath}

\author{Z. Nasr}
\affiliation{Department of Physics,
Faculty of Science, University of Isfahan,
Isfahan, Iran}
\author{F. Kheirandish}
\email{fkheirandish@yahoo.com}
\affiliation{Department of Physics,
Faculty of Science, University of Isfahan,
Isfahan, Iran}
\begin{abstract}
\noindent \textbf{Abstract:}
To remedy the failure of minimal coupling method in describing the quantum dynamics of two localized Brownian oscillators interacting with a common medium, a scheme is introduced to modeling the medium by a continuum of complex scalar fields or equivalently two independent real scalar fields. The starting point is a Lagrangian of the total system and quantization is achieved in the framework of canonical quantization. The equations of motion, memory or response functions and fluctuation-dissipation relations are obtained. An induced force between oscillators is obtained originating from the fluctuations of the medium. Ohmic regime and Drude regularization is discussed and the positions of oscillators are obtained approximately in large time limit and weak coupling regime.
\end{abstract}
\maketitle
\section{Introduction}
\noindent
Quantum dissipative systems are fundamental to many fields of physics: quantum optics \cite{Vogel, Gardiner,
Agarwal, Louisell}, super conductivity, quantum tunneling\cite{Caldeira1}, quantum information\cite{Vedral},
diffusion processes and quite generally open quantum systems \cite{Breuer}. In order to describe a dissipative
system quantum mechanically there exists two main approaches. In the first approach, the loss of energy is
phenomenologically introduced by modifying the Lagrangian or Hamiltonian of the system. In this approach usually
a fundamental feature of quantum mechanics,
for example the uncertainty relation is violated \cite{Caldirola,Kanai}. In the second approach, instead of modifying the Lagrangian or Hamiltonian of the main system, another system usually known as the heat bath or reservoir is considered and the total system known as system plus reservoir interact with each other usually in a linear way to have dissipation on energy or phase. In this approach, the dynamics of the system is described by Langevin equation \cite{Ford1,Ford2}. The reservoir is modeled by an infinite set of harmonic oscillators and a suitable coupling between the system and
reservoir is considered \cite{Caldeira2,Caldeira3,Caldeira4,Caldeira5}. Since the start point is a classical Lagrangian of the total system, the canonical quantization can be achieved systematically and consistently.

The total system can also be quantized in the frame work of the path integrals. In this case  effects of environmental degrees of freedom on the main system can be investigated using the Feynman-Vernon influence functional by tracing out environmental variables \cite{Feynman,Paz}.

In the present paper, following the second approach, the dynamics of two localized Brownian oscillators
embedded in a common medium is investigated. The minimal coupling method which has worked successfully in a
variety of problems \cite{Kheirandish1,Kheirandish2,Kheirandish3}, now suffers a major shortcoming for this
problem. Quite unexpectedly it leads to a non dissipative equation of motion for the relative distance between the
oscillators which is not physically acceptable in a dissipative medium. In order to remedy this deficiency, We have proposed a scheme in which the environment oscillators are modeled by a complex field consisting of two independent real scalar fields. This assumption leads to consistent results \cite{Duarte} and also provides us with a better understanding of the dynamics of the system.

The layout of the present work is as follows: In Sec. II, the model is introduced and starting from a Lagrangian the total system is quantized in the framework of canonical quantization approach. In S. III, the equations of motion are found, the memory or response functions are defined and fluctuation-dissipation relations are determined. In this section also an induced potential or force between oscillators is obtained originating from fluctuations of the medium. In Sec. IV, Ohmic damping and Drude regularization is discussed and the positions of oscillators are obtained approximately in large time limit and weak coupling regime. Finally, conclusions are given in Sec. V.
\section{The Model Lagrangian}
\noindent
The system which we are considering here consists of two one dimensional harmonic oscillators localized around the positions $x=0$ and $x=d$ respectively. The oscillators interact with a common medium causing an indirect interaction between oscillators. In order to investigate the quantum dynamics of the system we use the system plus reservoir method in which the heat bath is modeled by a continuum of complex fields $Y_{\nu}$ consisting of two independent real scalar fields
\begin{equation}\label{L1}
  Y_{\nu}=y_{\nu}+i z_{\nu},
\end{equation}
the Lagrangian of the whole system is
\begin{equation}\label{L2}
   L(t)=L_{R}+L_{s}+L_{int},
\end{equation}
where
\begin{equation}\label{L3}
  L_{s}=\frac{1}{2}M\dot{x}_{1}^{2}(t)+\frac{1}{2}M\dot{x}_{2}^{2}(t)-\frac{1}{2}M\omega_{0}^{2}x_{1}^{2}(t)-
  \frac{1}{2}M\omega_{0}^{2}x_{2}^{2}(t),
\end{equation}
is the lagrangian of the oscillators and
\begin{equation}\label{L4}
 L_{R}=\frac{1}{2}\int_{0}^{\infty}d\nu\,(\dot{Y}_{\nu}\dot{Y}_{\nu}^{*}-\omega_{\nu}^{2}Y_{\nu}Y_{\nu}^{*}),
\end{equation}
is the lagrangian of the heat bath. The dispersion relation may be taken arbitrarily but here we consider a linear dispersion relation $\omega_\nu =u_0 \,\nu$ where $u_0$ can be considered as the speed of propagation (sound) in the medium. If we rewrite this Lagrangian in terms of its constituent fields, we have
\begin{equation}\label{5}
 L_{R}=\frac{1}{2}\int_{0}^{\infty}d\nu\,(\dot{y}_{\nu}^{2}-\omega_{\nu}^{2}y_{\nu}^{2}
 +\dot{z}_{\nu}^{2}-\omega_{\nu}^{2}z_{\nu}^{2}).
\end{equation}
The interaction lagrangian is defined by
\begin{equation}\label{L5}
 L_{int}= \int_{0}^{\infty}d\nu\,(y_{\nu} \mbox{Re}[f_{\nu}(x_{1})+f_{\nu}(x_{2})]-z_{\nu}
 \mbox{Im}[f_{\nu}(x_{1})+f_{\nu}(x_{2})]),
\end{equation}
where the real and imaginary parts of the complex field are coupled with the real and imaginary parts of the
function $f_{\nu}(x)$ respectively.

The canonical conjugate variables corresponding to the dynamical variables of the system and the heat bath are
\begin{eqnarray}\label{conjugates}
p_{i}(t)&=&\frac{\p{L}}{\p{\dot{x}_{i}(t)}}= M \dot{x}_{i}(t),\,\,\,(i=1,2),\nonumber\\
p_{\nu}(t)&=&\frac{\p{L}}{\p{\dot{y}_{\nu}(t)}}=\dot{z}_{\nu}(t),\nonumber\\
q_{\nu}(t)&=&\frac{\p{L}}{\p{\dot{z}_{\nu}(t)}}=\dot{z}_{\nu}(t).
\end{eqnarray}
To quantize the whole system, the following equal-time commutators, among the variables and their conjugates,
should be imposed
\begin{eqnarray}
&& [\hat{p}_{i}(t),\hat{x}_{j}(t)] = i\hbar\,\delta_{ij},\,\,\,(i,j=1,2),\nonumber\\
&& [\hat{p}_{\nu}(t),\hat{y}_{\nu'}(t)] = i\hbar\,\delta (\nu-\nu'),\nonumber\\
&& [\hat{q}_{\nu}(t),\hat{z}_{\nu'}(t)] = i\hbar\,\delta(\nu-\nu').
\end{eqnarray}
Using definitions (\ref{conjugates}), Hamiltonian of the whole system is obtained as
\begin{eqnarray}
H &=&\frac{\hat{p}_{1}^2(t)}{2M}+\frac{\hat{p}_{2}^2(t)}{2M} \nonumber\\
&+&\frac{1}{2}\,\int_{0}^{\infty}d\nu\,\big[\hat{p}^2_{\nu}(t)+\omega^2_{\nu}\hat{y}^2_{\nu}(t)\big]+
\frac{1}{2}\,\int_{0}^{\infty}d\nu\,\big[\hat{q}^2_{\nu}(t)+\omega^2_{\nu}\hat{z}^2_{\nu}(t)\big]\nonumber\\
&+&\int_{0}^{\infty}d\nu\,\big\{\hat{z}_{\nu}(t)[f_{\nu}^{I}(\hat{x}_{1}(t))+f_{\nu}^{I}(\hat{x}_{2}(t))]-\hat{y}_{\nu}(t)[f_{\nu}^{R}(\hat{x}_{1}(t))+
f_{\nu}^{R}(\hat{x}_{2}(t))]\big\}.
\end{eqnarray}
Hamiltonian can also be rewritten in terms of creation and annihilation operators of the system and the heat bath which is more suitable for state transition calculations in the framework of Fermi's golden rule formula which we do not discuss it here. In order to do so we define the following ladder operators for the main system
\begin{equation}\label{L9}
\hat{a}_{j}(t)=\sqrt{\frac{M\omega_{0}}{2\hbar}}(\hat{x}_{j}+\frac{i}{M\omega_{0}}\hat{p}_{j}),\qquad
\hat{a}^{\dag}_{j}(t)=\sqrt{\frac{M\omega_{0}}{2\hbar}}(\hat{x}_{j}-\frac{i}{M\omega_{0}}\hat{p}_{j}),
\end{equation}
and the heat bath
\begin{eqnarray}
\hat{A}_{\nu}(t)&=&\sqrt{\frac{\omega_{\nu}}{2\hbar}}(\hat{y}_{\nu}+\frac{i}{\omega_{\nu}}\hat{p}_{\nu})\qquad
\hat{A}^{\dag}_{\nu}(t)=\sqrt{\frac{\omega_{\nu}}{2\hbar}}(\hat{y}_{\nu}-\frac{i}{\omega_{\nu}}\hat{p}_{\nu}),\nonumber\\
\hat{B}_{\nu}(t)&=&\sqrt{\frac{\omega_{\nu}}{2\hbar}}(\hat{z}_{\nu}+\frac{i}{\omega_{\nu}}\hat{q}_{\nu})\qquad
\hat{B}^{\dag}_{\nu}(t)=\sqrt{\frac{\omega_{\nu}}{2\hbar}}(\hat{z}_{\nu}-\frac{i}{\omega_{\nu}}\hat{q}_{\nu}).
\end{eqnarray}
Hamiltonian can now be rewritten as $ H=H_{0}+ H_{int}$,
\begin{eqnarray}
H_{0}&=&\hbar\omega_{0}(\hat{a}^{\dag}_{1}\hat{a}_{1}+\hat{a}^{\dag}_{2}\hat{a}_{2})
+\int_{0}^{\infty}d\nu\,\hbar\omega_{\nu}(\hat{A}^{\dag}_{\nu}\hat{A}_{\nu}+\hat{B}^{\dag}_{\nu}\hat{B}_{\nu}),\nonumber\\
H_{int}&=&\int_{0}^{\infty}d\nu\,g_{\nu}\,\sqrt{\frac{\hbar}{2\omega_{\nu}}}
\left\{(\hat{B}^{\dag}_{\nu}+\hat{B}_{\nu})\bigg[\sin(\nu\sqrt{\frac{\hbar}{2M\omega_{0}}}(\hat{a}_{1}+\hat{a}^{\dag}_{1}))
+\sin(\nu\sqrt{\frac{\hbar}{2M\omega_{0}}}(\hat{a}_{2}+\hat{a}^{\dag}_{2}))\bigg]\right\}\nonumber\\
&-&\int_{0}^{\infty}d\nu\,g_{\nu}\,\sqrt{\frac{\hbar}{2\omega_{\nu}}}\left\{(\hat{A}^{\dag}_{\nu}+\hat{A}_{\nu})
\bigg[\cos(\nu\sqrt{\frac{\hbar}{2M\omega_{0}}}(\hat{a}_{1}+\hat{a}^{\dag}_{1}))+
\cos(\nu\sqrt{\frac{\hbar}{2M\omega_{0}}}(\hat{a}_{2}+\hat{a}^{\dag}_{2})\bigg]\right\},
\end{eqnarray}
where in $H_{0}$, we have ignored from the irrelevant constant terms.
\section{Equations Of Motion}
\noindent
Using Heisenberg equations of motion we find the following set of coupled equations
\begin{align}\label{H1}
 M\ddot{{x}}_{i}(t)+M\omega^{2}_{0}x_{i}(t) &=
 \int_{0}^{\infty}d\nu\,\left(y_{\nu}(t)\frac{\p{f^{R}_{\nu}}(x_{i}(t))}{\p{x_{i}(t)}}-z_{\nu}(t)\frac{\p{f^{I}_{\nu}}(x_{i}(t))}{\p{x_{i}(t)}}\right),\\
 y_{\nu}(t)&=y_{\nu}^{N}(t)+\int_{0}^{t}dt'\,\frac{\sin[\omega_{\nu}(t-t')]}{\omega_{\nu}}[f_{\nu}^{R}(x_{1}(t'))+f_{\nu}^{R}(x_{2}(t'))],\\
z_{\nu}(t)&=z_{\nu}^{N}(t)-\int_{0}^{t}dt'\,\frac{\sin[\omega_{\nu}(t-t')]}{\omega_{\nu}}[f_{\nu}^{I}(x_{1}(t'))+f_{\nu}^{I}(x_{2}(t'))],
\end{align}
in the above equations, $ z_{\nu}^{N}(t)$ and $ y_{\nu}^{N}(t)$ are the solutions to the
homogenous equations of motion for the bath oscillators given by
\begin{align}\label{H2}
z_{\nu}^{N}(t)&=z_{\nu}^{N}(0)\cos \omega_{\nu}t +
 p_{\nu}^{N}(0)\frac{\sin\omega_{\nu}t}{\omega_{\nu}},\\
 y_{\nu}^{N}(t)&=y_{\nu}^{N}(0)\cos \omega_{\nu}t +
 q_{\nu}^{N}(0)\frac{\sin\omega_{\nu}t}{\omega_{\nu}},\\
\end{align}
where
\begin{eqnarray}
&& [z_\nu^N (0),p_{\nu'}^N (0)]=i\hbar\,\delta(\nu-\nu'),\nonumber\\
&& [y_\nu^N (0),q_{\nu'}^N (0)]=i\hbar\,\delta(\nu-\nu').\\
\end{eqnarray}
To proceed, choosing a suitable coupling function is necessary. Regarding the translational invariance of a
homogeneous medium we can assume \cite{Caldeira1}
\begin{equation}\label{H3}
 f_{\nu}(\mathbf{x}(t))= g_{\nu}\exp(i\nu x(t)),
\end{equation}
where $g_{\nu}$ is the coupling strength and $x(t)$ is the position of the oscillator with respect to a fixed point. In our case the position of the oscillators are $x_1 (t)$ and $d+x_2 (t)$ respectively. For this particular choice, the equations of motion for the oscillators are
\begin{align}\label{H4}
& M\ddot{x}_{1}(t)+M\omega^{2}_{0}x_{1}(t)
+M\int_{o}^{t}dt'\,\left[\chi(t-t';x_{1}(t)-x_{1}(t'))\dot{x}_{1}(t')+\chi(t-t';x_{1}(t)-x_{2}(t'))\dot{x}_{2}(t')\right]\notag\\
&+\frac{\p{V(x_{1}(t)-x_{2}(t))}}{\p{x_{1}(t)}}= M F_{1}(t),\nonumber\\
& M\ddot{x}_{2}(t)+M\omega^{2}_{0}x_{2}(t)
+M\int_{o}^{t}dt'\,\left[\chi(t-t';x_{2}(t)-x_{2}(t'))\dot{x}_{2}(t')+\chi(t-t';x_{1}(t)-x_{2}(t'))\dot{x}_{1}(t')\right]\notag\\
&+\frac{\p{V(x_{1}(t)-x_{2}(t))}}{\p{x_{2}(t)}}= M F_{2}(t),
\end{align}
where the response or memory functions $\chi(t-t';x_{i}(t)-x_{i}(t')),\,(i=1,2)$ are
defined by
\begin{eqnarray}\label{H5}
&& \chi(t-t';x_{i}(t)-x_{i}(t'))=
\int_{0}^{\infty}d\nu\,\frac{\nu^2 g_{\nu}^2}{\omega_{\nu}^2}\,\cos [\omega_{\nu}(t-t')]\cos [
\nu(x_{i}(t)-x_{i}(t'))],\nonumber\\
&& \chi(t-t';x_{1}(t)-x_{2}(t'))=
\int_{0}^{\infty}d\nu\,\frac{\nu^2 g_{\nu}^2}{\omega_{\nu}^2}\,\cos [\omega_{\nu}(t-t')]\cos [
\nu(x_{1}(t)-x_{2}(t')-d)].\nonumber\\
\end{eqnarray}
In Eqs. (\ref{H4}) an effective potential between oscillators is induced by the environment given by
\begin{equation}\label{H6}
 V(x_{1}(t)-x_{2}(t))= -\int_{0}^{\infty}d\nu\,\frac{g_{\nu}^2}{\omega_{\nu}^2}\cos[\nu(x_{1}(t)-x_{2}(t)-d)],
\end{equation}
this potential is absent when we use minimal model for quantization. The noise forces $ F_{i}(t)$ are defined
by
\begin{eqnarray}\label{H7}
&& F_{1}(t)=f_{1}(t)+\int_{0}^{\infty}d\nu\,\frac{\nu
g_{\nu}^2}{\omega_{\nu}^2}\left[\sin(\nu(x_{1}(t)-x_{1}(0)))+\sin(\nu(x_{1}(t)-x_{2}(0)-d))\right],\nonumber\\
&& F_{2}(t)=f_{2}(t)+\int_{0}^{\infty}d\nu\,\frac{\nu
g_{\nu}^2}{\omega_{\nu}^2}\left[\sin(\nu(x_{2}(t)-x_{2}(0)))+\sin(\nu(x_{2}(t)-x_{1}(0)+d))\right],
\end{eqnarray}
where
\begin{eqnarray}\label{H8}
&& f_{1}(t)=-\int_{0}^{\infty}d\nu\,\nu g_{\nu}\left[\mathbf{y}_{\nu}^{N}(t)\sin(\nu(x_{1}(t)))+\mathbf{z}_{\nu}^{N}(t)\cos(\nu(x_{1}(t)))\right],\nonumber\\
&& f_{2}(t)=-\int_{0}^{\infty}d\nu\,\nu g_{\nu}\left[\mathbf{y}_{\nu}^{N}(t)\sin(\nu(x_{2}(t)+d))+\mathbf{z}_{\nu}^{N}(t)\cos(\nu(x_{2}(t)+d))\right].
\end{eqnarray}
Using (\ref{H8}) and definitions (\ref{H5}), we find the fluctuation-dissipation relations
\begin{eqnarray}\label{FDR}
&& \langle f_i (t) f_i (t')\rangle_S=\int_0^{\infty}d\nu\,\frac{\nu^2\,g_\nu^2\,\hbar}{2\omega_\nu}\coth\big(\frac{\hbar\omega_\nu}{2K_B T}\big)
 \,\cos[\omega_\nu(t-t')]\,\cos[\nu(x_i (t)-x_i (t'))],\,\,\,(i=1,2),\nonumber\\
&& \langle f_1 (t) f_2 (t')\rangle_S=\int_0^{\infty}d\nu\,\frac{\nu^2\,g_\nu^2\,\hbar}{2\omega_\nu}\coth\big(\frac{\hbar\omega_\nu}{2K_B T}\big)
 \,\cos[\omega_\nu(t-t')]\,\cos[\nu(x_1 (t)-x_2 (t')-d)],\nonumber\\
\end{eqnarray}
where
\begin{equation}\label{sym}
 \langle f_i (t) f_j (t') \rangle_S=\langle \frac{f_i (t) f_j (t')+f_j (t')f_i (t)}{2}\rangle.
\end{equation}
In high temperature regime we recover
\begin{eqnarray}\label{FDRH}
&& \langle f_i (t) f_i (t')\rangle_S =K_B T\,\chi(t-t';x_{i}(t)-x_{i}(t')),\,\,\,(i=1,2),\nonumber\\
&& \langle f_1 (t) f_2 (t')\rangle_S =K_B T\,\chi(t-t';x_{1}(t)-x_{2}(t')).
\end{eqnarray}
To find more explicit relations let us assume the following form for the coupling function
\begin{equation}\label{C_form}
 g_\nu^2=A \,\nu^2\,e^{-\frac{\nu}{\nu_0}},
\end{equation}
where $A$ is a constant and $\nu_0$ may be considered as a cut off on wave number or frequency, that is the oscillators (main system) do not couple to bath oscillators with very high (also very low) frequencies or wave numbers. For this choice, we find for the mutual potential induced between oscillators
\begin{equation}\label{potential}
  V(u_{12})=-\frac{A\,\nu_0}{u_0^2}\,\frac{1}{1+\nu_0^2\,(u_{12}-d)^2},
\end{equation}
where $u_{12}=x_1 (t)-x_2 (t)$. Therefore, in such a medium the oscillators are attracted to each other with the force
\begin{equation}\label{force}
  F_{12}(u_{12})=-\frac{2A\,\nu_0^3}{u_0^2}\,\frac{u_{12}-d}{[1+\nu_0^2\,(u_{12}-d)^2]^2},
\end{equation}
From definitions (\ref{H5}) and the coupling strength (\ref{C_form}), we find for the memory functions
\begin{eqnarray}\label{memory}
&& \chi(t-t';u_{ii})=\frac{A\,\nu_0^3}{u_0^2}\,\bigg[\frac{1-3\nu_0^2\,u_{ij,+}^2}{(1+\nu_0^2\,u_{ii,+}^2)^3}+
 \frac{1-3\nu_0^2\,u_{ii,-}^2}{(1+\nu_0^2\,u_{ii,-}^2)^3}\bigg],\nonumber\\
&& \chi(t-t';u_{12})=\frac{A\,\nu_0^3}{u_0^2}\,\bigg[\frac{1-3\nu_0^2\,u_{12,+}^2}{(1+\nu_0^2\,u_{12,+}^2)^3}+
 \frac{1-3\nu_0^2\,u_{12,-}^2}{(1+\nu_0^2\,u_{12,-}^2)^3}\bigg],\nonumber\\
\end{eqnarray}
where for simplicity we have defined
\begin{eqnarray}\label{u}
&& u_{ii,\pm}=x_i (t)-x_i (t')\pm u_0 \,(t-t'),\nonumber\\
&& u_{12,\pm}=x_1 (t)-x_2 (t')-d\pm u_0 \,(t-t').
\end{eqnarray}

Since the oscillators are localized around the points $0$ and $d$, in the argument of response functions
(\ref{H5}) and the noise force (\ref{H7}) we can approximately set $ x_{1}(t)\approx {0}$, $
x_{2}(t)\approx {0}$, for all $t$. This approximation is valid if the cut off on frequency or wave number $\nu_0$ satisfies $\nu_0 \,a \ll 1$, where $a$ is the maximum of oscillator amplitudes. In this case, the induced force between oscillators will be cancelled by the second term in the noise forces (\ref{H7}) and we will find
\begin{align}\label{H9}
\ddot{x}_{1}(t)+\omega_{0}^2 x_{1}(t)+\int_{0}^{t}
dt'\,\left[\chi(t-t')\dot{x}_{1}(t')+\chi(t-t';d)\dot{x}_{2}(t')\right]=f_{1}(t),\nonumber\\
\ddot{x}_{2}(t)+\omega_{0}^2 x_{2}(t)+\int_{0}^{t}
dt'\,\left[\chi(t-t')\dot{x}_{2}(t')+\chi(t-t';d)\dot{x}_{1}(t')\right]=f_{2}(t),
\end{align}
where
\begin{equation}\label{H10}
\chi(t-t')=\int_{0}^{\infty}d\nu\,\frac{\nu^2 g_{\nu}^2}{\omega_{\nu}^2}\cos[\omega_{\nu}(t-t')],
\end{equation}
\begin{equation}\label{H11}
\chi(t-t';d)=\int_{0}^{\infty}d\nu\,\frac{\nu^2 g_{\nu}^2}{\omega_{\nu}^2}\cos[\omega_{\nu}(t-t')]\cos(\nu d).
\end{equation}
Equation (\ref{H10}) is a cosine transform and from its inverse we find
\begin{equation}\label{CF}
  g_\nu^2=\frac{2\omega_\nu^2}{\pi\nu^2}\,\int_0^\infty dt\,\chi(t)\,\cos(\nu t),
\end{equation}
so the coupling strength $g_\nu$ can be determined if the memory function $\chi (t)$ is given. To decouple the equations (\ref{H9}) we define the new variables
\begin{eqnarray}
 R(t) &=& \frac{x_1 (t)+x_2 (t)}{2},\nonumber \\
  Z(t) &=& x_2 (t)-x_1 (t),
\end{eqnarray}
note that these variables are not center of mass and relative distance of oscillators since $x_1$ and $x_2$ are defined as oscillations around the equilibrium positions $x=0$ and $x=d$, respectively. In terms of these variables we have
\begin{eqnarray}\label{EQS}
\ddot{R}(t)+\omega_{0}^{2}R(t)+\int_{0}^{\infty}dt'\,\chi_R(t-t';d)\,\dot{R}(t')&=&F_{R}(t),\nonumber\\
\ddot{Z}(t)+\omega_{0}^{2}Z(t)+\int_{0}^{\infty}dt'\,\chi_Z(t-t';d)\,\dot{Z}(t') &=& F_{Z}(t),
\end{eqnarray}
where for simplicity we have defined
\begin{eqnarray}\label{defs}
  F_{R}(t) &=& (f_1 (t)+f_2 (t))/2, \nonumber\\
 F_{Z}(t) &=& f_{2}(t)-f_{1}(t),\nonumber\\
  \chi_R(t-t';d) &=& \chi (t-t')+\chi (t-t',d)=
  \int_{0}^{\infty}d\nu\,\frac{2\nu^2 g_{\nu}^2}{\omega_{\nu}^2}\cos[\omega_{\nu}(t-t')]\cos^2(\nu d/2),\nonumber \\
  \chi_Z(t-t';d) &=& \chi (t-t')-\chi (t-t',d)=
  \int_{0}^{\infty}d\nu\,\frac{2\nu^2 g_{\nu}^2}{\omega_{\nu}^2}\cos[\omega_{\nu}(t-t')]\sin^2(\nu d/2).
\end{eqnarray}
The dissipation-fluctuation relations for the new variables in high temperature regime are
\begin{eqnarray}
\langle F_R (t) F_R (t') \rangle_S &=& \frac{1}{2}K_B T \, \chi_R(t-t';d), \\
\langle F_Z (t) F_Z (t') \rangle_S &=& 2 K_B T \, \chi_Z(t-t';d).
\end{eqnarray}
Equations (\ref{EQS}) can be solved formally using Laplace transform technique as
\begin{eqnarray}
R(t) &=& \xi_{+}(t)R(0)+\eta_{+}(t)\dot{R}(0)+\int_{0}^{t}dt' \eta_{+}(t')F_{R}(t-t'),\nonumber\\
Z(t) &=& \xi_{-}(t)Z(0)+\eta_{-}(t)\dot{Z}(0)+\int_{0}^{t}dt' \eta_{-}(t')F_{Z}(t-t'),
\end{eqnarray}
where
\begin{eqnarray}
\eta_{\pm}(t) &=& L^{-1}\left[\frac{1}{s^{2}+s[\widetilde{\chi}(s)\pm \widetilde{\chi}(s,d)]+\omega_{0}^{2}}\right] \qquad \xi_{\pm}(t)=L^{-1}\left[\frac{s+\tilde{\chi}(s)\pm\tilde{\chi}(s,d)}{s^{2}+s[\widetilde{\chi}(s)\pm \widetilde{\chi}(s,d)]+\omega_{0}^{2}}\right],
\end{eqnarray}
these equations for physical memory functions are complicated to be solved analytically and numerical calculations should be applied. An interesting feature about equations (\ref{EQS}) is that we can not choose both memory functions $\chi_R$ and $\chi_Z$ to be memory less or Markovian, since the response functions $\chi_R$ and $\chi_Z$ are not independent. One can find $\chi (t:d)$ in terms of $\chi (t)$ from definitions (\ref{H10}, \ref{H11}) easily.

\section{Ohmic Damping}
\noindent
In response functions (\ref{H9}), let us assume the following coupling
\begin{equation}\label{C1}
 g_\nu^2=\frac{u_0^3 \gamma}{\pi},
\end{equation}
a more physical coupling function will be considered in the following. The corresponding memory functions are
\begin{equation}\label{C2}
 \chi (t-t')=\gamma\,\delta(t-t'),\,\,\,(t>t'),
\end{equation}
\begin{equation}\label{C3}
 \chi (t-t';d)=\frac{\gamma}{2}\,\delta(t-t'-d/u_0),\,\,\, (t>t').
\end{equation}
Inserting the memory functions (\ref{C2}, \ref{C3}) into equations (\ref{H9}) we find
\begin{equation}\label{C4}
\ddot{x}_{1}(t)+\omega_{0}^{2}\,x_{1}(t)+\frac{\gamma}{2}\,\dot{x}_{1}(t)
= f_{1} (t)-\frac{\gamma}{2}\,\dot{x}_{2}(t-\tau_0) ,
\end{equation}
\begin{equation}\label{C5}
\ddot{x}_{2}(t)+\omega_{0}^{2}\,x_{2}(t)+\frac{\gamma}{2}\,\dot{x}_{2}(t)
= f_{2}(t)-\frac{\gamma}{2}\,\dot{x}_{1}(t-\tau_0) ,
\end{equation}
where we have defined a characteristic time $\tau_0 =d/u_0 $ and made use of $\int_0^t dt'\,\delta(t-t')=1/2$, since $t$ is on the border of the interval $[0,t]$. From equations (\ref{C4}, \ref{C5}) it is seen that the oscillators feel each other after a retardation time $\tau_0$ as expected. For weak coupling regime $\gamma\ll\omega_0 $, we can solve equations (\ref{C4}, \ref{C5}) as series in $\gamma$, in large time regime we find
\begin{eqnarray}\label{X12}
  x_1 (t) &=& \int_0^t dt'\,G(t-t')\,f_1 (t')-\frac{\gamma}{2}\,\int_0^t dt'\,\int_0^{t'} dt''\,\frac{\partial}{\partial t'}\,G(t'-t'')\,f_2 (t''),\nonumber\\
  x_2 (t) &=& \int_0^t dt'\,G(t-t')\,f_2 (t')-\frac{\gamma}{2}\,\int_0^t dt'\,\int_0^{t'} dt''\,\frac{\partial}{\partial t'}\,G(t'-t'')\,f_1 (t''),
\end{eqnarray}
where
\begin{equation}\label{Green}
 G(t-t')=\Theta(t-t')\,e^{-\gamma (t-t')/4}\,\frac{\sin[\Omega\,(t-t')/4]}{\Omega/4},
\end{equation}
is the Green's function of the damped harmonic oscillator and $\Theta (t)$ is the Heaviside step function. From equations (\ref{X12}) one can find correlation functions $\langle x_i (t)\,x_j (t')\rangle $ using fluctuation-dissipation relations (\ref{FDR}). As we already mentioned, ohmic case is an idealized situation in real physical world, there is always a microscopic time scale for inertia effects in the reservoir \cite{Weiss}. The simplest generalization is to regularize the damping kernel to the so-called Drude regularization
\begin{equation}\label{Drude}
  \chi(t)=\gamma\omega_{D}\Theta(t)\exp(-\omega_{D}t),
\end{equation}
for which $\tau_{D}=\omega_{D}^{-1}$ is Drude memory time. In the limit of $\omega_{0}\ll\omega_{D}$, the reservoir behaves like an ohmic one. With this choice, using (\ref{CF}) the coupling strength becomes
\begin{equation}\label{nonohmic}
  g_{\nu}^2 = \frac{2\gamma u_{0}^3}{\pi}\,\frac{1}{1+ \lambda_{D}^2 \nu^{2}},
\end{equation}
where $\lambda_D=u_0 /\omega_D$. The memory function $\chi(t;d)$ can also be obtained as
\begin{equation}
  \chi(t;d)=\frac{\gamma \omega_D}{2}\,[e^{-\omega_D |t+d/u_0|}+e^{-\omega_D |t-d/u_0|}],
\end{equation}
note that when $d\rightarrow 0$, we recover (\ref{Drude}), also in the limit $d\rightarrow\infty$, we have $\chi(t;d)\rightarrow 0$, that is oscillators are not correlated in this case as expected.
\section{Conclusion}
\noindent
To remedy the deficiency of minimal coupling method in describing two independent quantum particles interacting with a common medium, a scheme is introduced which is consisting with the previous methods. In this scheme medium or heat-bath is modeled by a continuum of complex fields or equivalently two independent real scalar fields. A Lagrangian for the total system is introduced and quantization is achieved in the framework of canonical quantization approach. An induced force between the particles of the main system is obtained which is originated from the fluctuations of the medium. The memory or response functions are introduced and fluctuation-dissipation relations are obtained. The Ohmic regime and Drude's regularization is discussed and approximate positions of localised oscillators are obtained in large time limit for weak coupling regime.

\end{document}